\documentclass[11pt]{article}
\usepackage{aaspp4}
\def\~{\roma{\sim}}
\def\roma#1{\ifmmode{#1}\else{$#1$}\fi}

\def\Rqt{\roma{\rm R^{1/4}}}

\def\etal{{\it et al.}}

\def\kms{\roma{\,\rm km\,s^{-1}\,}}           
\def\kmsMpc{\roma{\rm\,km\,s^{-1}\,Mpc^{-1}}} 
\def\kmsmpc{\roma{\rm\,km\,s^{-1}\,Mpc^{-1}}}

\def\Dn{\roma{\rm D_n}}
\def\Dk{\roma{\rm D_K}}

\def\Dv{\roma{\rm D_V}}

\def\Ho{\roma{\rm H_o}}

\def\dsig{\roma{\rm D-\sigma}}

\slugcomment{Accepted for publication in {\em New Astronomy}}
\begin{document}

\title {\bf The Coma -- Leo~I Distance Ratio and the Hubble Constant}

\author{Michael~D.~Gregg
}
\affil{Institute for Geophysics and Planetary Physics\\
Lawrence Livermore National Laboratory\\
L-413, 7000 East Avenue\\
Livermore, CA 94550\\
gregg@igpp.llnl.gov}

\begin{abstract}

The diameter - velocity dispersion relation in B, V, and K for three
early type galaxies in the Leo~I (M96) group is derived from published
photometry and kinematic data.  The relations in all three colors have
slopes which agree well with those for the Coma cluster.  The RMS
scatter of the Leo~I galaxies in each color is extremely small,
consistent with the group's compactness.  These relations yield
estimates of the Coma--Leo~I distance ratio of $9.01 \pm 0.51$, $8.77
\pm 0.43$, and $8.82 \pm 0.31$ respectively, with a weighted mean of
$8.84 \pm 0.23$.  The general agreement among the three colors indicates
that the early-type galaxies in Leo~I and Coma have similar stellar
populations.

The Coma--Leo~I distance ratio coupled with estimates of the absolute
distance to the Leo~I group allows the Hubble constant to be
determined, free of the uncertainties which arise when working with
the Virgo cluster.  Several high quality distance estimates are
available from a variety of techniques: Cepheids in M96 (Tanvir \etal\
1995) and M95 (Graham \etal\ 1997), surface brightness
fluctuations (Tonry \etal\ 1997), planetary nebulae luminosity
functions (Ciardullo \etal\ 1993), and the luminosity of the red giant
branch tip (Sakai, Freedman, \& Madore 1996).  Adopting a cosmic
recession velocity of the Coma cluster in the microwave background
frame of $7200 \pm 300$ \kms, these distance estimates lead to values
of the Hubble constant ranging from 70 to 81 \kmsmpc, with an
unweighted mean of $75 \pm 6$ \kmsmpc.

\end{abstract}

\pagebreak

\section {Introduction}

Cepheid distances to at least five Virgo cluster galaxies have now
been obtained using the Hubble Space Telescope (see van den Bergh 1996
for a summary).  While these studies are milestones in the effort to
establish the extragalactic distance scale, they have not yet settled
the controversy over the value of the Hubble constant for two critical
reasons.  First, it is now widely recognized that the line-of-sight
depth of the Virgo cluster is large, perhaps as great as a factor of
1.5 or more (Tonry, Ajhar, \& Luppino 1990; Tammann \etal\ 1996).  The
Hubble constant based on even many Virgo spirals will be subject to
uncertainty because the centroid of the Cepheids, in spiral galaxies,
relative to the centroid of the Virgo cluster as measured by
elliptical galaxies is not well determined.  Second, both the infall
of the Milky Way into the Virgo cluster and the cosmic recession
velocity of Virgo are still controversial.  Estimates range from 100 to
450 km/s, large compared to the mean Virgo recession velocity of $\sim
1100$ km/s (Huchra 1988; Tammann \etal\ 1996).

The uncertainty in distance introduced by the depth of the Virgo
cluster can be overcome by working with more compact objects, such as
the Fornax cluster or the Leo~I group, an approach being taken by the Hubble
Key Project on the Extragalactic Distance Scale
(http://www.ipac.caltech.edu/H0kp) and others (see Freedman 1996
for a summary).  The additional uncertainty introduced by the local
peculiar velocity field remains a problem for distance scale
determinations using clusters within a few thousand \kms,
where peculiar velocities may be a significant fraction of the Hubble
velocity.  It can be mitigated by determination of the Hubble
constant, not directly from the recession velocity of the nearby
cluster, but by using the relative distance to the Coma cluster to
step well beyond the local velocity field.  This technique has been
used by the HST Extragalactic Distance Scale Key Project (Farrarese
\etal\ 1996) and by Pierce \etal\ (1994) to derive the Hubble
constant from Cepheid measurements in Virgo and by Tanvir \etal\
(1995) in the Leo~I group.

The Leo~I group is relatively nearby, with a mean recession velocity
of $\sim 850 \kms$, and is compact, with a line-of-sight depth
estimated to be just $\sim 2\%$ compared to its distance, assuming
spherical symmetry.  It contains two spirals, NGC3368 (M96) and
NGC3351 (M95), which now have Cepheid distances (Tanvir \etal\ 1995;
Graham \etal\ 1997) and several early type galaxies, including
NGC~3377 (E6), NGC~3379 (E0), and NGC~3384 (SB0).  In principle, use
of the Leo~I -- Coma distance ratio plus a direct measure of the
distance to Leo~I allows the Hubble constant to be determined free of
the uncertainties introduced by the Milky Way's space motion and the
large depth of Virgo.

The present work derives the Coma -- Leo~I distance ratio based on
photometrically independent diameter-velocity dispersion relations in
B, V, and K bands using data in the literature.  The internal
uncertainties are small, establishing the relative distance with
precision as $8.84 \pm 0.23$.  Adoption of a distance to the Leo~I
group allows the Hubble constant to be determined without reference to
the Virgo cluster.  Five Leo~I group distances with small errors have
been published and range from 10.05 Mpc (Graham \etal\ 1997) to 11.6
Mpc (Tanvir \etal\ 1995); curiously, the Cepheid distances define the
two extremes.  These imply a Hubble constant in the range $70 \pm 7$
to $81 \pm 8$ \kmsmpc.

\section{D-$\sigma$ Relations}

The \dsig\ relation for elliptical galaxies is based on a near-optimal
projection of the fundamental plane (Dressler \etal\ 1987; Djorgovski
\& Davis 1987; Jorgensen, Franx, \& Kjaergaard 1993).  The diameter D
is defined as the size of a circular aperture which encloses a
fiducial mean surface brightness: 20.75 in the B band (Dressler \etal\
1987), 19.80 in V (Lucey \etal\ 1991), or 16.75 in K (Gregg 1995a).
The diameters, referred to as \Dn, \Dv, and \Dk, respectively, are
tightly correlated with central velocity dispersion ($\sigma$) and
provide a relative distance indicator for early-type galaxies of
accuracy comparable to and perhaps better than the infrared
Tully-Fisher relation for spirals (Dressler \etal\ 1987; Gregg 1995a).

\subsection{Coma and Virgo clusters}

The K-band \dsig\ relation for 24 E and S0 galaxies in the Coma
cluster and 12 E galaxies in Virgo was derived by Gregg (1995a) from
the IR photometry of Bower, Lucey, \& Ellis (1992a,b) and Persson,
Frogel, \& Aaronson (PFA, 1978).  The velocity dispersions come
primarily from Davies \etal\ (1987).  Figure~1 shows the K band \dsig\
relations from Gregg (1995a) for Coma and Virgo along with the B and V
band relations from Faber \etal\ (1989) and Bower, \etal\ (1992a,b).
The \dsig\ relation for the Coma cluster is significantly tighter in
the K band than in either B or V (Gregg 1995a).  The RMS dispersion of
the 24 Coma galaxies in B, V, and K is 0.035, 0.029, and 0.021 in
log(D) and 8\%, 7\%, and 5\% in distance, respectively.  The lines
through the Coma \dsig\ relations are derived using a robust fitting
technique (see Gregg 1995a for details); the lines through the Virgo
data are forced to have the same slopes as the Coma fits and the
residuals are minimized in a least squares sense by choice of
intercepts.

The large scatter in all colors exhibited by the Virgo cluster can be
attributed to its depth along the line of sight, providing graphical
illustration of the difficulties in determining its centroid for
either absolute or relative distance estimates.  The envelope
containing most of the Virgo galaxies in these diagrams implies a
front-to-back distance spread of $\sim 1.5$ {\em for the elliptical
galaxies in Virgo}, similar to the spread found for Virgo spirals and
implying that not even the early-type galaxies in Virgo define a
locally condensed cluster core.  Some of this spread can perhaps be
attributed to aperture effects in determining the central velocity
dispersion (Jacoby 1995), but much of it is probably real.

\subsection{Leo~I}

PFA also give multiaperture K and V band data for three Leo~I early
type galaxies, NGC~3377, 3379, and 3384.  The diameters \Dk\ and \Dv\
can be derived by fitting an \Rqt\ law (Gregg 1995a).  Burstein \etal\
(1987) and Davies \etal\ (1987) give B-band \Dn\ and velocity
dispersion data for the ellipticals, but not for NGC~3384.  To include
this galaxy in the analysis requires adopting the velocity dispersion
of McElroy (1995), corrected for its redshift by Equation~1 of Davies
\etal\ (a 5\% effect).  The B-band diameter, \Dn, for NGC~3384 has
been estimated using its B-V of 0.95 (de Vaucouleurs \etal\ 1991) and
the V-band photometry from PFA; as such, this one diameter is not
independent of the V-band data.  Leaving NGC~3384 out of the B-band
relation does not significantly affect the results.  For comparison
and as a test of the fitting technique, B-band diameters (called
D$_{\rm B}$ to distinguish them from published D$_{\rm n}$)
for NGC~3377 and NGC~3379 were derived from the PFA data, adopting B-V
colors from Burstein \etal\ (1987).  In both cases, the derived diameters
agree to much better than 0.01 in log(D) as given in Burstein \etal.

Table~1 lists the diameters and adopted velocity dispersions for
the 3 Leo~I galaxies along with other relevant data.  The reddening
model used by PFA predicts no reddening for the Leo~I group.  Their V
and K photometry have been corrected here using the Galactic reddening
estimates of Burstein \& Heiles (BH, 1984).

The resulting B, V, and K \dsig\ relations for the three Leo~I
galaxies are shown in Figure~1.  All display remarkably little
scatter, particularly the K-band.  The dashed lines have the same
slopes as derived for the Coma cluster fits; the intercepts are
determined by minimizing the residuals.  The RMS residuals in log(D)
for Leo~I for B, V, and K are 0.029, 0.018, and 0.009, corresponding
to a scatter in distance of only 6.9\%, 4.2\%, and 2.1\%,
respectively; the K-band relation is consistent with the estimated
depth of the cluster along the line of sight.  The very small scatter
obtained for the Leo~I galaxies is probably fortuitous as the
estimated errors for the velocity dispersions by themselves should
produce a somewhat larger scatter, especially at K.  To be
conservative in estimating the total uncertainties, it is assumed that
the error per point in Leo~I is equivalent to the RMS scatter as
determined empirically from the larger sample in the Coma \dsig\
relations.

The derived intercepts for B, V, and K are 0.900, 0.995, and 1.298,
respectively.  Combined with the Coma cluster fits (Gregg 1995a),
these imply $9.01 \pm 0.51$, $8.77 \pm 0.43$, and $8.82 \pm 0.31$ for
the Coma/Leo~I distance ratio.  The uncertainty in the relative
distance is the uncertainty in the separation of the fits to the two
clusters and can be estimated as
\begin {equation} 
\sum_{Coma} {\sigma_{i} \over \sqrt N} + \sum_{Leo} {\sigma_{i} \over
\sqrt M}
\end {equation}
where N and M are the number of points in Coma (24) and in Leo~I (3).  The
three estimates agree within the uncertainties at considerably better than
the $1\sigma$ level, suggesting that the adopted errors are, indeed,
conservative. The weighted mean distance ratio is $8.84 \pm 0.23$.  This
direct measure of the distance ratio agrees remarkably well with the Tanvir
\etal\ (1995) value obtained by a more circuitous route through Virgo
based on many different distance indicators: $8.85
\pm 0.32$.

\subsection{Systematics Affecting in the Coma--Leo~I Distance Ratio}

There are at least two possible systematic effects which may influence
the Coma--Leo~I distance ratio as derived from the \dsig\ relation:
reddening and stellar populations.  Neither appears to be significant.

\subsubsection{Stellar Populations}

If a systematic difference in age, metallicity, or initial mass function
(IMF) exists between the stellar populations of the early type galaxies in
Coma and Leo~I, then the slopes and/or zeropoints of the \dsig\ relations
would be different in the two environments and the derived relative
distance would be in error (Gregg 1995b).  Systematic age differences in
early type populations are expected to be in the sense that objects in
dense environments have greater mean ages than those in small groups,
suggesting that the Leo~I galaxies would be younger than the Coma objects.
Younger, brighter ellipticals in Leo~I would cause them to appear nearer,
relative to Coma, and the Coma/Leo~I distance ratio would be overestimated.
A value of \Ho\ based on such an affected distance ratio would be a {\it
lower} limit.  Systematic effects due to stellar population variations will
be a function of wavelength (Gregg 1995b), being greater at B than V or K.
The consistency of the distance ratios derived in the three bandpasses
argues that the stellar populations do not differ greatly in age, metal
abundance, or initial mass function at a given velocity dispersion between
the two environments.  The similarities of the slopes of the \dsig\
relations in Leo~I to those derived for Coma also imply a consistency in
the stellar populations in the two environments.

Figure~2a plots V-K (PFA, BLE) colors against velocity dispersion for
the Leo~I (open circles), Coma (solid dots) and Virgo (filled
diamonds) early-type galaxies.  BLE corrected the V-K colors for Virgo
and Coma to remove aperture effects due to color gradients to
facilitate comparison; the colors are for aperture sizes of 60\arcsec\
and 11\arcsec, respectively, which were chosen to be equal to the
relative distance of the two clusters.  The PFA V-K colors for Leo~I
are for aperture an aperture size of 56\arcsec; to be completely
compatible with the BLE data, these should be corrected to an aperture
of 90\arcsec.  While the PFA V-K data for Leo~I do exhibit color
gradients of $\sim 0.05-0.1$ between 29\arcsec\ and 56\arcsec,
correction to a 90\arcsec\ aperture will be only a few hundredths of a
magnitude bluer at most and will not alter the main conclusion from
Figure~2a, that the Leo~I galaxies do not have unusual V-K colors.
The lower panels of Figure~2 demonstrate that the Leo~I galaxies have
typical B-V (Burstein \etal\ 1987) and Mg$_2$ indices (Davies \etal\
1987; no Mg$_2$ is available for NGC~3384).  The conclusion again is
that the Leo~I early-type objects have normal stellar populations.

\subsubsection{Reddening}

The normal colors of the Leo~I ellipticals (Figure~2) can also be
interpreted as evidence that the Galactic reddening is well determined
and that any internal reddening must be extremely small.  The
agreement of the distance ratios derived in the three photometric
bands B, V, and K also indicates that the reddening is correct.

\section{The Distance to the Leo~I Group and the Hubble Constant}

Using the Coma--Leo~I distance ratio, a value for the Hubble constant
can be derived by adopting an absolute distance to the Leo~I group and
a cosmic recession velocity for Coma in the microwave background
frame.  Estimates of the Coma recession velocity are generally in good
agreement at approximately 7200 \kms\ (Faber \etal\ 1989; van den
Bergh 1996).  Following Tanvir \etal\ (1995), an uncertainty of 300
\kms\ is adopted to allow for the possibility of a sizable peculiar
velocity for the entire Coma cluster.

There are currently 4 methods which give distances to the Leo~I group
with quoted errors of 10\% or less: Cepheids (to two spirals, M96 and
M95), surface brightness fluctuations (SBF), planetary nebulae
luminosity functions (PNLF), and the location of the red giant branch
tip luminosity.  The distances and resultant Hubble constants are
summarized in Table~2; distance estimates to the Coma cluster are also
included.  The latter values amount to zeropoint calibrations of the
\dsig\ relations of the Coma cluster based on the other indicators.
The error from the relative Coma--Leo~I distance is combined in
quadrature with the published error for each method to give the final
uncertainties listed in Table~2.

\noindent {\em Surface Brightness Fluctuations}

Tonry \etal\ (1997) present a compilation of
recalibrated SBF distance estimates to various nearby groups and
clusters.  The SBF Leo~I group distance is $10.67 \pm 0.30$ Mpc, based
on observations of 5 member galaxies: the 3 early-type galaxies used
in the \dsig\ relation plus another S0, NGC~3412, and the bulge of
M96.  This implies a value of $\Ho = 76 \pm 4 \kmsmpc$.

\noindent {\em Planetary Nebulae Luminosity Function}

Ciardullo, Jacoby, \& Tonry (1993) list PNLF distances to the the 3 Leo~I
early-type galaxies used here to derive the distance ratio with Coma.  The
weighted mean is $10.33 \pm 0.30$ Mpc, implying $\Ho = 79 \pm 5 \kmsmpc$.
This is consistent at the $1\sigma$ level with the SBF result.  Ciardullo
\etal\ (1993) point out that SBF and PNLF distances can always be brought
into agreement by adjusting the reddening since reddening affects the
distances derived by the two techniques in opposite ways.  For the Leo~I
early type galaxies, the small difference can be eliminated if
A$_{\rm V}$ is reduced to approximately zero; given the good agreement,
however, and other possible sources of error, this is perhaps best taken as
additional evidence that the Galactic reddening towards Leo~I
is indeed low.


\noindent {\em RGB Tip Luminosity}

Sakai \etal\ (1996) have detected the tip of the red giant branch in
NGC3379 from deep F814W (I-band) images from HST.  This yields a distance
of $11.4 \pm 0.8$ Mpc to Leo~I, implying a distance of $101 \pm 8$ Mpc
to Coma and \Ho = $71 \pm 6$ kmsMpc.

\noindent {\em Cepheids in M96 and M95}

Based on observations of 7 Cepheids in the HST F555W (V) and F814W (I)
bands, Tanvir \etal\ (1995) derive a distance to M96 of $11.6 \pm
0.9$ Mpc, leading to a Hubble constant of $72 \pm 7$.  Graham \etal\
(1997) report discovery of 49 Cepheids in NGC~3351, also using HST
F555W and F814W images, deriving a distance of $10.05 \pm 0.88$ Mpc.
Combined with the Coma-Leo~I distance ratio of 8.84, this yields \Ho =
$81 \pm 8$.

\noindent {\em Other Distance Estimators}

The distance to Leo~I has also been estimated using the globular cluster
luminosity function approach (Harris 1990), based on a combined sample
of clusters from NGC~3377 and 3379, as $10.7 \pm 2.2$ Mpc.  The
uncertainty is large because of the small number of globulars found and
more than spans the total range of more precise distance estimates,
and so provides no additional distance information.  

The Tully-Fisher (TF) method can be applied to M96, but its
inclination is $47\arcdeg$, making the distance uncertain by $\sim 1$
magnitude (Willick 1996, private communication).  Even with a
more favorable inclination, the TF uncertainty would still be sizable;
various estimates for the TF distance to M96 range from 11.0 Mpc
(Bottinelli \etal\ 1985), to 14.1 (Federspiel \etal\ 1996), both from
B-band photometry, with uncertainties approaching 20\% magnitudes.  An
H-band TF distance can be derived for M96 based on the re-calibration
of the extensive Aaronson database that has been carried out by
Tormen \& Burstein (1995) and Willick \etal\ (1997).  The present
author's approximate absolute calibration of the Willick \etal\ data
set results in a distance modulus of $11.0 \pm 1.7$.  These distances
and error bars span the total range of the other estimates for
Leo~I and, once again, there is no additional distance information.

\section{Discussion}

The above various distance estimates to Leo~I and resultant values for
the Hubble constant 
are illustrated graphically 
in Figure~3.  The right
ordinate scale displays the implied Hubble constant for each distance,
adopting a Coma -- Leo~I distance ratio of 8.84.  The error bars
indicate the published uncertainties for each of the distances; these
are not determined consistently from one to another.  The Cepheid
distance estimates have attempted to include known possible systematic
effects and hence their error bars are larger.

Perhaps somewhat unexpectedly, the two Cepheid distance estimates
bracket the other distance estimates to Leo~I.  The other three
techniques derive distances directly to the early-type galaxies used
in the \dsig\ relations; however, using the Cepheid distances to M95
and M96 to calibrate the \dsig\ relations relies on the assumption
that the spirals are at the same distance as the early-type members, a
possible source of systematic error.  Based on detailed dynamical
analysis and modeling of the intergalactic hydrogen ring in the Leo~I
group, Schneider (1989) makes strong arguments in favor of the
early-type galaxies NGC3377, NGC3379, NGC3384, and NGC3412, and the
spirals M96 and M95 being in close physical proximity, probably all
within 0.5~Mpc or less of each other.  Schneider's data show that M96,
in particular, appears to be directly interacting with the HI which is
orbiting the close pair NGC~3379 and 3384.  M95 is $1\fdg5$ away from
the NGC~3379/3384 pair; the evidence linking it to the early-type
galaxies is more tenuous.  But if M95 is really 1.3 Mpc closer than
M96, as the Cepheid results indicate, then the Leo~I group is extended
along the line-of-sight by 15\%, pointing almost directly towards us.

From Schneider's analysis, however, it is unlikely that both M96 and
M95 could be at distances much different than the early-type galaxies
in Leo~I.  The spread in the above cited distances determinations to
Leo~I can be interpreted simply as the real uncertainty in present
distance estimates.  Since possible systematic effects are difficult
to evaluate, and may enter in common in some of the estimators,
perhaps the best estimate of the Hubble constant in this instance is a
straight unweighted mean of all the methods (the heavy horizontal line
in Figure~3).  This yields a distance to Leo~I of $10.8 \pm 0.7$ Mpc,
to Coma of $96 \pm 7$ Mpc, and a Hubble constant of $75 \pm 6$
\kmsMpc, where the uncertainty is the dispersion of the various
estimates about the mean.  Because the Cepheid distances of Tanvir et
al.\ (1995) and Graham \etal\ (1997) bracket the range of distance
estimates, consideration of the Cepheids alone does not change this
result at all.

This result is additional support for the ``short'' distance scale,
consistent with a number of other studies, in particular the HST Key
Project on the distance scale (Freedman 1996).  If the Hubble constant
is instead close to 50 \kmsMpc, then one or more systematic errors
must exist in the arguments or in the data used here.  For \Ho\ to be
less than 60 \kmsMpc, then the Leo~I early-type galaxies would have to
be at a distance of $\sim 14.5$ Mpc, but this is $> 3\sigma$ greater
than either the Cepheid distance to M96 (Tanvir \etal\ 1995) or the
RGB tip distance to NGC3379 (Sakai \etal\ 1996), the two largest
accurate distance estimates to Leo~I.  Placing both M96 and M95 3-5
Mpc in the foreground of the early-type galaxies in Leo~I requires
setting aside the dynamical evidence of Schneider (1989) for a direct
physical link between M96 and NGC3379 and also requires that all three
of the early-type distance estimates to Leo~I (SBF, PNLF, and RGB tip)
are seriously in error at the 30-40\% level.  This must be considered
unlikely.  The Hubble constant may also be overestimated here if the
adopted velocity of the Coma cluster were seriously in error.  The
mean velocity of the Coma cluster is observationally well established,
but if the Coma cluster has a large peculiar velocity directed away
from us, then \Ho\ would be lower than estimated here.  To bring it
below 60 \kmsMpc, however, would require a peculiar velocity for Coma
of $\sim 1500$ \kms, which must be considered highly unlikely; most
studies of peculiar velocity fields yield estimates of a few hundred
\kms and Faber \etal\ (1989) estimate a peculiar velocity for Coma of
only 259 \kms, directed {\em towards} us.  Finally, the Hubble constant
could be pushed as low as 60 \kmsMpc if the Coma--Leo~I relative
distance were $\sim 11$ rather than $8.84 \pm 0.23$ as found here.  It
is argued above in \S 2.3 that this is unlikely; the most likely
systematic effect, a younger age for the early-type galaxies in Coma
and Leo~I, would work in the opposite direction, causing the relative
distance to be overestimated and \Ho\ underestimated.

\section{Summary}

The main result of this work is that the Coma--Leo~I distance ratio is
$8.84 \pm 0.23$ based on the \dsig\ relations in B, V, and K for three
early type galaxies in the Leo~I.

Consideration of the five most precise distance estimates to the Leo~I
group allows a distance to the Coma cluster to be derived and from
that an estimate of the Hubble constant which is free of the problems
associated with working in the Virgo cluster.  The distance estimates
used are based on Cepheids, SBF, PNLF, and RGB tip estimates and span
a range from 10.05 to 11.6 Mpc, somewhat surprisingly with the two
Cepheid determinations to M96 and M95 bracketing the range.  Utilizing
the Coma--Leo~I distance ratio derived here and adopting a cosmic
recession velocity for the Coma cluster of 7200 km/s, the
corresponding Hubble constants range from $70 \pm 7$ to $81 \pm 8
\kmsMpc$ with an unweighted mean of $\Ho = 75 \pm 6 \kmsMpc$.

\noindent {\em Note added in proof:}

In a paper in press, Hjorth \& Tanvir (1997) derive the Coma -- Leo I
distance ratio based on the optical fundamental plane for 5 early type
galaxies in Leo I, NGC3412 and NGC3489 in addition to the three used
in this paper.  Their results of $9.12 \pm 0.67$ and $9.51 \pm 0.67$ for
the distance ratio based on angular diameters and luminosities,
respectively, agree within $1 \sigma$ with the result presented here,
$8.84 \pm 0.23$, using the diameter - velocity dispersion approach.

\acknowledgments

This research has made use of the NASA/IPAC Extragalactic Database
(NED) which is operated by the Jet Propulsion Laboratory, Caltech,
under contract with the National Aeronautics and Space Administration.
This work was done at the Institute of Geophysics and Planetary
Physics, under the auspices of the U.S. Department of Energy by
Lawrence Livermore National Laboratory under contract
No.~W-7405-Eng-48.


\pagebreak 

\pagebreak

%
\begin{deluxetable}{lccccccccc}
\tablenum{1}
\tablewidth{0pt}
\tablecaption{Data for Leo~I Early-Type Galaxies}
\tablehead {
\colhead {NGC} &
\colhead {Type} &
\colhead {A$_{\rm B}$} &
\colhead {Log($\sigma$)} &
\colhead {Log(D$_{\rm n}$)} &
\colhead {Log(D$_{\rm B}$)} &
\colhead {Log(D$_{\rm V}$)} &
\colhead {Log(D$_{\rm K}$)} &
\colhead {(B-V)$_0$} &
\colhead {(V-K)$_0$}
}
\startdata
N3377 & E6 & 0.06 & 2.116 & 1.838 & 1.835 & 1.816 & 1.796 & 0.90 & 2.99 \nl
N3379 & E0 & 0.05 & 2.303 & 2.018 & 2.019 & 2.027 & 2.050 & 0.97 & 3.27 \nl
N3384 & SB0& 0.07 & 2.216 &\nodata& 1.929 & 1.923 & 1.935 & 0.95 & 3.13 \nl
\enddata
\end{deluxetable}
\begin{deluxetable}{lcccl}
\tablenum{2}
\tablewidth{0pt}
\tablecaption{Distances to Leo~I and Coma and the Implied Hubble Constant}
\tablehead {
\colhead {Method} & 
\colhead {D Leo} &
\colhead {D Coma} &
\colhead{{\rm H$_{\rm o}$}} &
\colhead{Reference} \\
\colhead {} &
\colhead {(Mpc)} &
\colhead{(Mpc)} & 
\colhead{(km s$^{-1}$ Mpc$^{-1}$)} &
\colhead {}
} 
\startdata
Cepheids M95   &  $10.05 \pm 0.88$ & $\phn88.8 \pm 8.1$ & $81 \pm 8$ & Graham et al. (1997) \nl
PNLF           &  $10.33 \pm 0.30$ & $\phn91.3 \pm 3.6$ & $79 \pm 5$ & Ciardullo et al. (1993) \nl
SBF            &  $10.67 \pm 0.30$ & $\phn94.3 \pm 3.6$ & $76 \pm 4$ & Tonry et al. (1996) \nl
RGB tip        &  $11.4\phn \pm 0.8\phn$   & $100.8 \pm 7.5$& $71 \pm 6$  & Sakai et al. (1996) \nl
Cepheids M96   &  $11.6\phn \pm 0.8\phn$   & $102.5 \pm 7.6$& $70 \pm 6$ & Tanvir et al. (1995) \nl
Unweighted mean& $10.8\phn \pm 0.7\phn$    & $\phn95.5 \pm 6\phn$   & $75 \pm 6$  & ~ \nl
\enddata
\end{deluxetable}

\begin {figure}
\plotfiddle{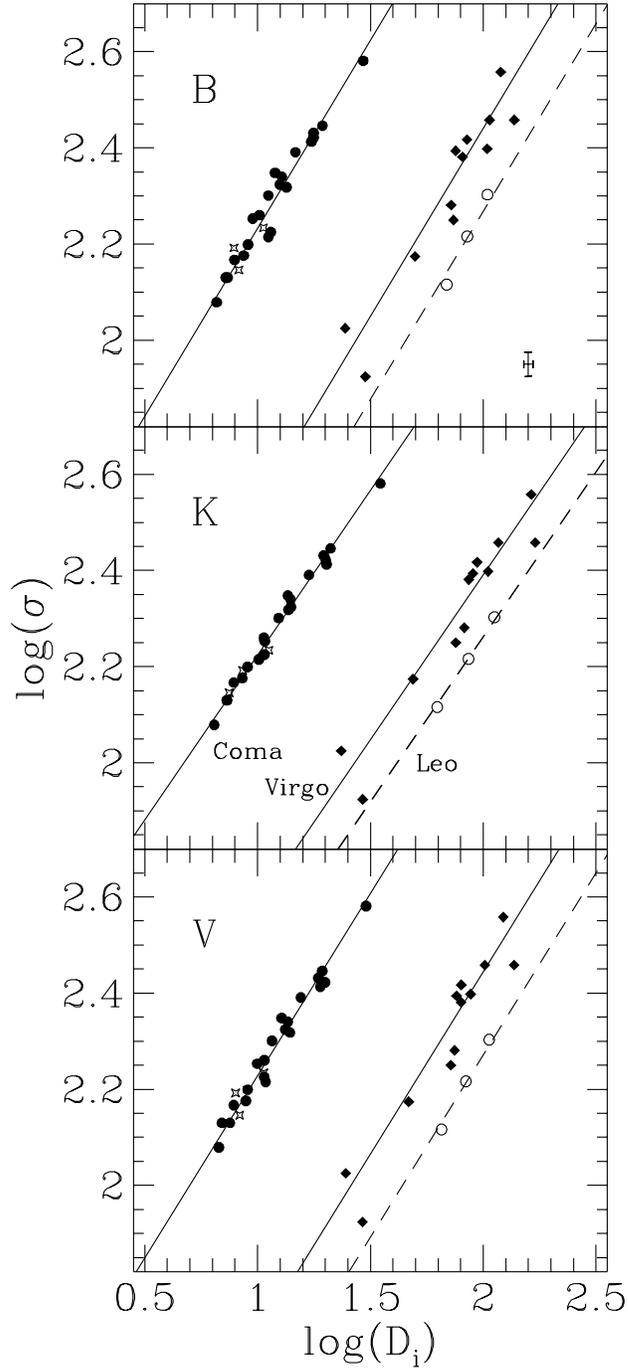}{7in}{0}{85}{85}{-240}{-70}
\caption {Comparison of the B, V, and K \dsig\ relations for the Coma
and Virgo clusters and the Leo~I group.  Filled circles are Coma
ellipticals, 4-pointed stars are Coma S0's, filled diamonds are Virgo
ellipticals, and open circles are the Leo~I early-type galaxies.  The
lines through the Coma cluster data are robust fits.  The lines
through the Virgo and Leo~I data sets are forced to have the same
slope as the Coma relations and the fits are achieved by adjusting
only the intercept.  The Leo~I relations have slopes in excellent
agreement with those for Coma and the Leo~I galaxies exhibit very
little scatter.}
\end {figure}

\begin {figure}
\plotone{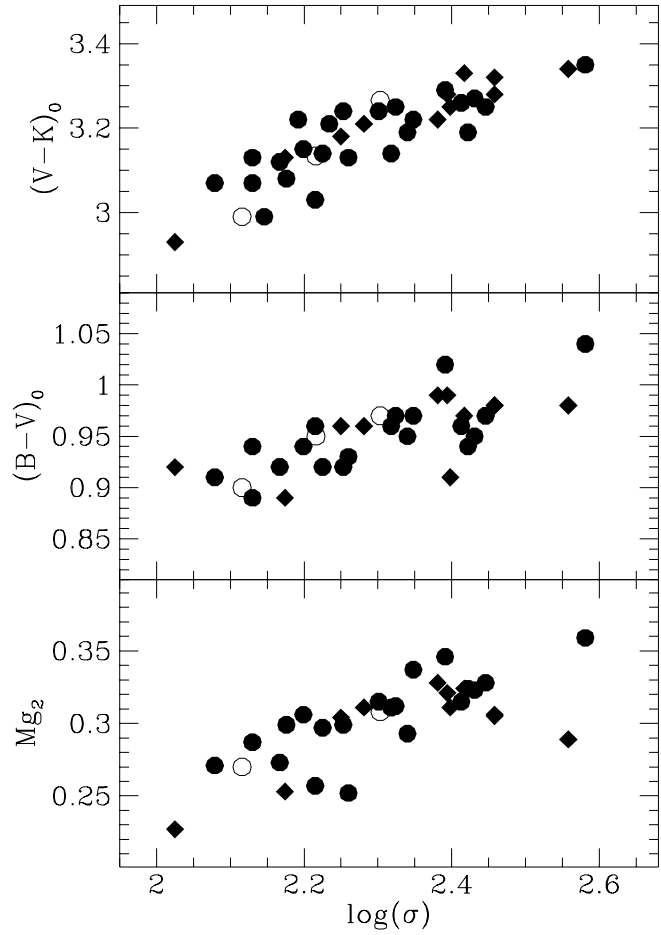}
\caption {Comparisons of V-K, B-V, and Mg$_2$ as a function of
log($\sigma$) for Coma, Virgo, and the Leo~I early-type galaxies.
Symbols as in Figure~1.  The Leo~I galaxies have typical colors and
Mg$_2$ strength compared to ellipticals in the other clusters.}
\end {figure}

\begin {figure}
\plotfiddle{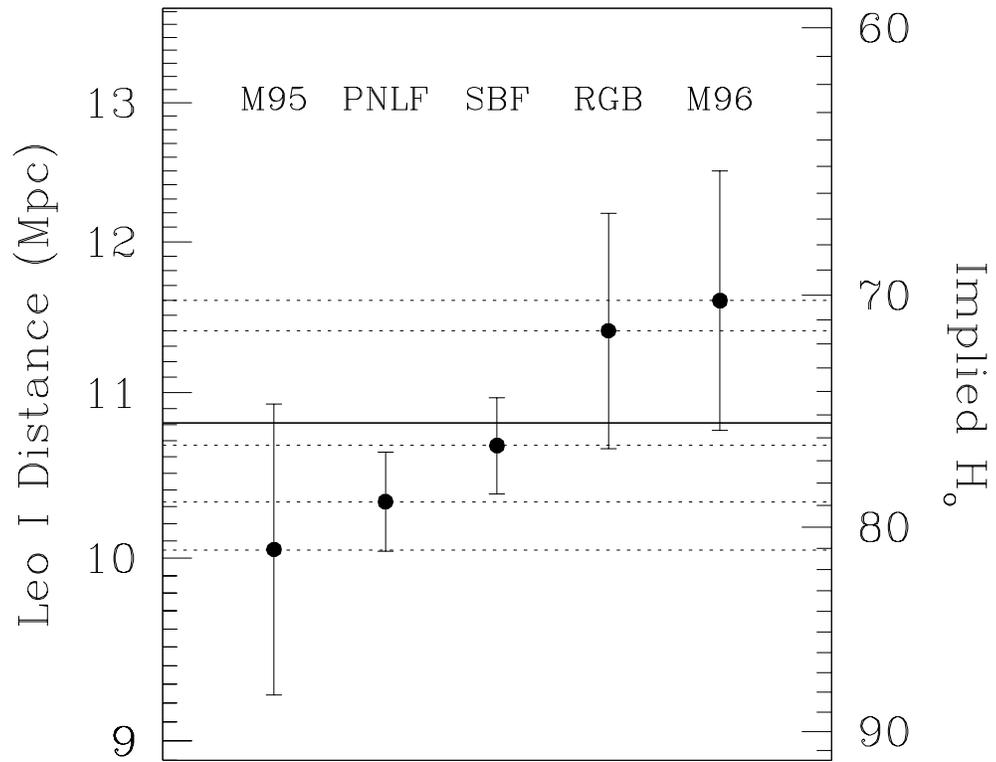}{4in}{0}{90}{90}{-330}{-230}
\caption {Graphical comparison of the best available distance
estimates to the Leo~I group and the implied values of the Hubble
constant using the Coma--Leo~I distance ratio of $8.84 \pm 0.23$ from
$\S 2.2$.  The points labeled M96 and M95 are Cepheid distances from
Tanvir \etal\ (1995) and Graham \etal\ (1997), respectively.}
\end {figure}

\end{document}